\documentclass[aps,prx,preprint,groupedaddress]{revtex4-2}

\usepackage{amssymb}
\usepackage{amsmath}
\usepackage{amsthm}
\usepackage{mathtools}
\usepackage{amsfonts}
\usepackage{xcolor}
\newtheorem{Proposal}{\bf Proposal}[section]

\numberwithin{equation}{section}
\begin{document}

\title{Mechanism of wavefunction collapse in measurements of separated quantum subsystems}
\author{Gregory D. Scholes}
\affiliation{Department of Chemistry, Princeton University, Princeton, NJ 08544, USA}
\date{\today}
%%%% Keyword entries to be placed here %%%%
%\keywords{nonlocality, collapse, entanglement, quantum mechanics}

%%%% Insert corresponding author and its email address}
%\corres{Gregory D. Scholes\\
%\email{gscholes@princeton.edu}}

%\date{\today}

\begin{abstract}
The specific advance of this work is to propose a mechanism by which superpositions collapse during measurement of the separated subsystems of entangled quantum states.  It is shown how the phase that locks together entangled states could play a special role in the measurement of isolated subsystems. This `contextual' phase is installed into separated subsystems  of an entangled state,  and decides the measurement outcomes for the subsystems by directing the collapse of each superposition to a particular classical outcome when a subsystem is measured. The measuring apparatus thus obtains a classical read-out of the quantum correlations embedded in an entangled state.  More broadly, these results suggest insights into the theory of measurement of quantum superpositions.  
\end{abstract}

\maketitle
\newpage
%%%%%%%%%%%%%%%%%%%%%%%%%%%%
%\tableofcontents

\section{Introduction}

When a light beam is incident on a 50-50 beam splitter, we observe that half the intensity is transmitted and half is reflected. We can perform the same experiment one photon at a time\cite{KnightQOptics}. Then we observe that half the time, randomly, the incident photon is transmitted and half the time it is reflected. The theory of quantum mechanics predicts this statistical outcome, but an open question is what mechanism determines whether the photon will be transmitted or reflected? The physical outcome is prescribed by the projection postulate of quantum mechanics that says a superposition state must `collapse' when measured to yield a physical observable. This postulate is not debated in practice, but it suggests a perplexing open question:  what is a plausible mechanism to explain state collapse of isolated systems within the linear theory of quantum mechanics? Proposing a possible answer to this question is the focus of the present paper.

To formulate the question clearly, let's consider a pair of entangled photons produced by a nonlinear process so their polarizations are in the prototypical singlet state:
\begin{equation}
	\Psi_- = \frac{1}{\sqrt{2}} \Big( |H\rangle_A |V\rangle_B - |V\rangle_A |H\rangle_B \Big) .
\end{equation}
where A and B label the optical modes, which are spatially separated after being produced in the entangled state. H and V label polarization as horizontal and vertical with respect to a particular axis system. What is perplexing is the notion that, prior to any measurement, each photon in the entangled pair exists as a superposition of alternatives (each could be polarized H or V in the detection frame). If we examine $N$ entangled photon pairs, performing measurements on photon A through a polarizer set to V, we find that $N/2$ photons are detected. A similar result is found if we change the polarizer to H. That is consistent with the quantum-mechanical prediction from the reduced density matrix for subsystem A, which is $\rho_A = \text{Tr}_B |\Psi_-\rangle \langle \Psi_- | = \frac{1}{2} I_2$, where $I_2$ means the $2 \times 2$ identity matrix. This is a mixed state, indicating that photon A is equally likely to be H- or V-polarized. 

This outcome is explained in the Copenhagen interpretation, or projection postulate of quantum theory, to result from a `collapse' of the superposition state accompanying measurement\cite{Laloe, GriffithsBook, OmnesBook}. However, it is recognized that a mechanism for collapse under unitary evolution of the wavefunction is unknown and has reasonably been criticized as being `\emph{ad hoc}'\cite{Dickson}. Nevertheless, collapse of some kind seems to be needed  so that measurements always have outcomes\cite{AlbertVaidman}.

We can, further, compare measurements jointly recorded for photon A and photon B. In the $z-$basis, we find that if photon A is H-polarized, then photon B will be V-polarized, and vice versa. That anticorrelation is predicted by finding an expectation of $-1$ for the joint operator $\sigma_z \otimes \sigma_z$ acting on $\Psi_-$ in the tensor product space\cite{Peres} $\mathcal{H}_A \otimes \mathcal{H}_B$. Here $\sigma_z$ is a Pauli operator. However, what is not known is precisely how the separated photon polarizations become anticorrelated through the mechanism of random collapse of the relevant superpositions. 

The paper thereby discusses an issue that has troubled researchers for decades\cite{Mermin1985, Laloe, OmnesBook, Peres, Ballentine, CS1978, GHSZ1990, ReidEPR, hidden10, unspeakable}. As noted in the Einstein, Podolsky and Rosen (EPR) paper\cite{EPR}, what we really would like to explain is that while entangled states assign ``two different wave functions to the same reality'', when measurement of one subsystem reveals its outcome, how is it that measurement of the other subsystem---well separated spatially from the first---finds perfect correlation with the first subsystem's measurement outcome without any kind of unlikely interaction happening faster than the speed of light? This apparent aspect of nonlocality is especially troubling\cite{Vaidman2019, Zwirn2023}.

The issue is widely considered closed because the possible explanations, mainly based on hidden local variables, have been ruled out by experimental demonstrations that quantum systems can violate Bell's inequalities\cite{Aspect2015}.  However, that reasoning has been shown to need revision\cite{Brassard}. The present paper reports that there exists a mechanism explaining how superpositions collapse,  which concomitantly resolves the question of why measurement outcomes on separated quantum subsystems of an entangled state are correlated.

The main aim of the present paper is to suggest a way to explain---entirely within the theory of linear quantum mechanics---the mechanism of collapse and its implications for the joint measurements on separated subsystems. To clarify why our approach is relevant to the measurement problem, and, in particular, why we focus on entangled states rather than superposition states of a single quantum system, we give some background\cite{OmnesBook}. Consider a quantum system prepared in a superposition of basis states:
\begin{equation*}
	\psi_A = \frac{1}{\sqrt{2}} (|0\rangle_A + |1\rangle_A).
\end{equation*} 

If we make a measurement in the $\{ |0\rangle_A , |1\rangle_A \}$ basis, then according to the quantum formalism we obtain one measurement outcome, either $|0\rangle_A$ or $|1\rangle_A$, each equally probable. Thus, one way of looking at the problem of how definite measurement outcomes are obtained is to focus on collapse of the superposition state $\psi_A$. That is the viewpoint suggested by Schrödinger's famous cat analogy. 

However, there is a second way to approach this measurement problem. That is, with explicit account for the measuring device. Let's consider the interaction of the quantum system in state $\psi_A$ with a measuring apparatus\cite{OmnesBook}. According to the Schödinger equation, this interaction generates an entangled state between the quantum system and measuring device, for example:
\begin{equation*}
	\Psi_{AM} = \frac{1}{\sqrt{2}} ( |0\rangle_A |0\rangle_M + |1\rangle_A |1\rangle_M ).
\end{equation*}

Now the problem of measurement outcomes is to explain how this \emph{entangled} state `collapses' into one outcome. For example, the quantum system collapses to $|0\rangle_A$ and the measurement device concomitantly records $|0\rangle_M$. Thus, the problem of how a superposition of a single quantum system yields a definite measurement outcome can be viewed equivalently  as the problem of how $\Psi_{AM}$ collapses to give a definite measurement reading---meaning a definite outcome for subsystem A and for subsystem M. For a concrete example, consider the measurement of the polarization state of a single photon in a polarization superposition (see Sec. 2 of \cite{EPR3}).

In the present paper, we therefore consider the general case of how definite measurement outcomes are obtained from entangled states within the linear theory of quantum mechanics. Thus, the second open question addressed by this paper is how measurements of the \emph{separated} photons (subsystems, more generally) yield the observed correlation.

\section{Measurements on separated subsystems}

Entangled states comprise intriguing superpositions of states associated with the constituent subsystems that allow for correlations in measurements exceeding those possible for analogous classical systems. For example, we can test various polarization combinations for joint measurements on an entangled photon pair. In terms of operators on $\Psi_-$, possible polarization combinations include polarization angles of H and V (giving the correlation predicted by the operator  $\sigma_z \otimes \sigma_z$) and with detection polarizers set at 45$^{\circ}$ to H and V (giving the correlation predicted by the operator $\sigma_x \otimes \sigma_x$)\cite{Mermin1993, Peres}, or any intermediate angle. The fact that quantum systems do not have all these measurement outcomes predetermined is exemplified by the Bell test, Kochen-Specker theorem and its variations\cite{Mermin1993, Peres}. An important result of such tests is the conclusion that correlations in the outcomes of measurements of separated subsystems cannot be explained by some kind of hidden classical variables that decide the observations. The correlations evidently defy classical interpretation, and are said to be nonlocal.  

Here an analysis of this problem is described that is based on measurements on separated, entangled subsystems. The subsystems are assumed to be well-separated, then the measurement process for subsystem A is independent of that for B. See, for example, the discussion in ref \cite{OmnesBook}. To account for this separation, we posit that the measurement apparatus obtains an expectation value for vectors in the local Hilbert space of a subsystem, $\mathcal{H}_A$ or $\mathcal{H}_B$, in order to map the entangled quantum state to classical indicators. 

\begin{Proposal}
	(Principle of separated subsystems) If two subsystems of an entangled state are separated sufficiently so that each can be characterized using independent measurements, then correlations  produced by entanglement between the subsystems can be predicted by considering measurements in the Hilbert space of each subsystem. 
\end{Proposal}

This concept is generally accepted\cite{vonNeumann}. What is distinct here from the common understanding is to explain why we cannot predict the relevant measurement outcomes for the separated subsystems using the linear operator acting on states in the tensor product space, that is, using  $\langle \Psi_- | O_A \otimes \mathbb{I}_B | \Psi_- \rangle$, for some operator $O_A$ acting on subsystem A. Instead, we need to project these measurement outcomes from the more general state space that maps to the tensor product space via the usual quotient map (to be described in Sec. 4). This new approach allows canonical projections from an entangled state to states associated with each subsystem. Importantly, the projection is well defined, not only for separable states\cite{vonNeumann}, but also for entangled states. According to this perspective, measurement is a map from $\mathcal{H}_A \otimes \mathcal{H}_B$ to $\mathcal{H}_A$ and $\mathcal{H}_B$ such that classical measurement instruments act by collapsing the subsystem superpositions in their respective Hilbert spaces that are projected from the overarching space (the free vector space). The proposal is justified further in Sec. 5, and supported by physical examples. 

To support the proposition, three consistency arguments are now given. The first consistency argument is that a measurement on a separated subsystem should give an equivalent outcome whether a second subsystem, entangled to the first, is present or whether it has been destroyed before the measurement. Clearly, if the subsystem B has been destroyed, then a subsequent measurement of the state of subsystem A  will yield a definite measurement outcome of the wavefunction of A in $\mathcal{H}_A$. That is, a measurement outcome of one subsystem will be consistent whether the subsystem is entangled with a second subsystem, or the second subsystem has been destroyed after entanglement.

For an entangled state $\Psi_-$, measurement of subsystem A in $\mathcal{H}_A \otimes \mathcal{H}_B$ gives the expectation value calculated from $\langle \Psi_- | O_A \otimes \mathbb{I}_B | \Psi_- \rangle$, for some operator $O_A$ that acts only on subsystem A. Here $\mathbb{I}_B $ is the identity operator on subsystem B.  In this case, the wavefunction does not collapse (in the absence of decoherence)  when a measurement is performed in this way on associated subsystems. The difference between measurements on the composite system and genuinely separated subsystems is discussed further in Sec. 5. We show why a linear operator like $O_A \otimes \mathbb{I}_B$ cannot resolve he expectation value that gives definite measurement outcomes for $\psi_A \in \mathcal{H}_A$. See also Sec. 3 of \cite{EPR3} and Sec. 6 below.

The \emph{probabilistic outcome} of measurements in $\mathcal{H}_A \otimes \mathcal{H}_B$, predicted by tracing over the density matrix $| \Psi_-\rangle \langle \Psi_- |$ with respect to subsystem B, gives a mixed state $\text{Tr}_B |\Psi_-\rangle \langle \Psi_- | = \frac{1}{2} I_2$, as mentioned above and as is well known. That is the same prediction as found for  an ensemble of measurements of subsystems A in $\mathcal{H}_A$, as is evident from analysis  described below. This is the second consistency test for Proposal 2.1. 

For the entangled photon example, measurement of the state in a subsystem's local Hilbert space means that the local vectors are added at the amplitude level. That is implemented physically by the detection polarizer. This suggests a third consistency test for Proposal 2.1, whereby the outcome of a classical measurement of a superposition of light polarization states should give the same result as a measurement on  separated quantum system that is in a polarization state superposition.

\begin{figure}
	\includegraphics[width=6.5 cm]{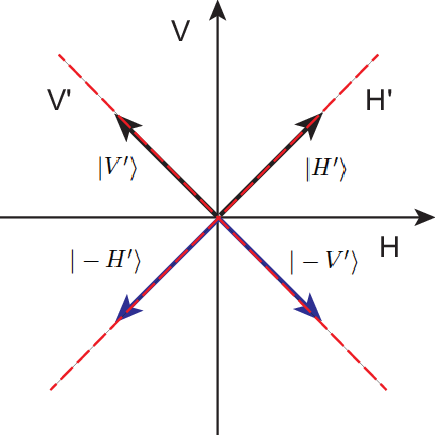}
	\caption{Polarization relationships. The H-V axis showing the settings for the $H'$ and $V'$ polarizers (dashed lines). The two polarization vectors and their antipodal partners are indicated by the bold arrows. }
	\label{fig1}
\end{figure}   

As an example of how measurements of separated subsystems work, according to Proposal 2.1, we can rewrite Eq. 1.1 in an equivalent form that explicitly shows the result of 45$^{\circ}$ measurements. Define the polarizer settings, shown in Fig. 1, as $|H'\rangle = \frac{1}{\sqrt{2}}(|H \rangle + |V \rangle)$ and $|V'\rangle = \frac{1}{\sqrt{2}}(|H \rangle - |V \rangle)$. Further, the wavefunction will be written in a way that absorbs the phase of the superposition (the minus sign in the case of Eq. 1.1) into the state vector of one of the subsystems, so we will also use $|-H'\rangle = \frac{1}{\sqrt{2}}(-|H \rangle - |V \rangle)$ and $|-V'\rangle = \frac{1}{\sqrt{2}}(-|H \rangle + |V \rangle)$. Two examples of equivalent versions of $\Psi_-$, written in the $x-$basis (in a form convenient for seeing correlations in measurements using $\sigma_z \otimes \sigma_z$ detection), are therefore
\begin{eqnarray}
	\Psi_-(\text{class 1}) &= \frac{1}{\sqrt{2}} \Big[ |V'\rangle_A \otimes |H'\rangle_B + |H'\rangle_A \otimes |-V'\rangle_B  \Big] \\
	\Psi_-(\text{class 2}) &= \frac{1}{\sqrt{2}} \Big[ |V'\rangle_A \otimes |H'\rangle_B + |-H'\rangle_A \otimes |V'\rangle_B  \Big] .
\end{eqnarray}
In the $z-$basis (convenient for seeing correlations in measurements using $\sigma_x \otimes \sigma_x$ detection) we have
\begin{eqnarray}
	\Psi_-(\text{class 1}) &= \frac{1}{\sqrt{2}} \Big[ |H\rangle_A \otimes |V\rangle_B + |-V\rangle_A \otimes |H\rangle_B  \Big] \\
	\Psi_-(\text{class 2}) &= \frac{1}{\sqrt{2}} \Big[ |H\rangle_A \otimes |V\rangle_B + |V\rangle_A \otimes |-H\rangle_B  \Big] .
\end{eqnarray}

By expanding each version of $\Psi_-$ and collecting terms, it is seen that Eqs 1.1, 2.1--2.4 are identical.

We will say more about the phase factors now. A key point to note is that if we consider real coefficients for linear combinations of the basis states in each subsystem, then we only need to consider phases that give signs $\pm$, as shown in the examples above. However, we can generalize that. Consider the entangled states, labeled $s$ for a state comprising a symmetric linear combination of simple tensors and $a$ for a state comprising an antisymmetric linear combination of simple tensors:
\begin{equation*}
	\Psi_s = \frac{1}{\sqrt{2}} (u_A \otimes u_B + w_A \otimes w_B)
\end{equation*}
and 
\begin{equation*}
	\Psi_a = \frac{1}{\sqrt{2}} (u_A \otimes u_B - w_A \otimes w_B).
\end{equation*}
We can write either of these states in more general form as
\begin{equation}
	\Psi = \frac{1}{\sqrt{2}} (e^{i\xi_A + i\xi_B} u_A \otimes u_B + e^{i\phi_A + i\phi_B} w_A \otimes w_B),
\end{equation}
where $\xi_A, \xi_B, \phi_A$, and $\phi_B$ are arbitrary phases, such that $\xi_A$ is the global phase of $u_A$, $\phi_A$ is the global phase of $w_A$, etc. We require for the symmetric linear combination $\Psi_s$ that
\begin{equation}
	|\xi_A + \xi_B  - \phi_A - \phi_B| = 2n\pi
\end{equation}
and for the antisymmetric state $\Psi_a$ that
\begin{equation}
	|\xi_A + \xi_B  - \phi_A - \phi_B| = (2n+1)\pi ,
\end{equation}

where $n \in \mathbb{Z}$. There are infinitely many phase combinations for any state, but it turns out that we can divide these possibilities into two unique classes when there are two subsystems (A and B). The proof is as follows. Any maximally entangled state comprising two subsystems can be written up to a global phase for the state as 
\begin{equation*}
	\Psi = \frac{1}{\sqrt{2}} (u_A \otimes u_B + \lambda (w_A \otimes w_B)),
\end{equation*}
where $\lambda$ defines the phase of the superposition: $\lambda = e^{\phi_A + \phi_B  - \xi_A - \xi_B}$. Without loss of generality, we can rotate the arbitrary phases of the first term so that $\xi_A = \xi_B  = 0$, then focus on the phases of the second term. For a specific example consider a symmetric superposition with $\lambda = 1$. The same reasoning can then be extended to any other superposition. Since we require $\phi_A + \phi_B = 0$, we can choose phase pairs such as $\phi_A = \phi_B = 0$, $\phi_A = \frac{\pi}{4}$ and $\phi_B = -\frac{\pi}{4}$, and so on. In practice, the choice is irrelevant except when we perform measurements on the separated subsystems. In that case, any $\phi_A + \phi_B = 0$ with $\phi_A, \phi_B \in (-\frac{\pi}{2}, \frac{\pi}{2}]$ defines a particular measurement configuration. Notice that each $\phi_A$ (and $\phi_B$) has a corresponding antipodal vector in the opposite half of $\mathbb{S}^1$, that is for values $\phi_A, \phi_B \in (-\pi, -\frac{\pi}{2}] \cup (\frac{\pi}{2}, \pi]$. Since these vectors differ only by an overall phase, they cannot be discriminated by measurements---but they give complementary outcomes for measurements on the separated subsystems. Each vector has precisely one corresponding antipodal vector, and hence there are two equivalence classes of contextual phase.

The two phase classes are implied by conditions (2.6) and (2.7). Let's consider the case when $\lambda$ is real. Even multiples of $\pi$ mean that the equivalence classes of phases are represented by $e^{\phi_A}$ and $e^{\phi_B}$ are both equal to 1 or are both $-1$. Odd multiples of $\pi$ mean that the equivalence classes of phases are represented by one phase factor having the value 1 and the other $-1$.

The class 1 and class 2 versions of $\Psi_-$ are identical in the space $\mathcal{H}_A \otimes \mathcal{H}_B$ because, by construction, maps on vectors in the tensor product space are linear\cite{TensorSpaces}. The way phase in the superpositions is identified with subsystem A or B, therefore, is irrelevant for the composite states. This is easily seen by noting that the 45$^{\circ}$ polarizer settings cannot distinguish antipodal vectors, Fig. 1, for example, $|H'\rangle $ from $|-H'\rangle $. However, the identification turns out to matter when the two subsystems are separated and measurements are performed on the states in $\mathcal{H}_A$ and $\mathcal{H}_B$. 

In the sense that measurement outcomes depend on these phases, the phases provide a form of \emph{context} associated to the state. Contextuality, in the Copenhagen interpretation of quantum mechanics means that the experimental outcome of a measurement depends on the measurement setup\cite{Jaeger}. Grangier elaborates on this concept\cite{Grangier2021, hidden9, CSM}. by proposing that objective properties---properties belonging to the quantum object---are defined for a particular measurement context. The phases described in the present work provide context additional to, and closely related to, the measurement basis\cite{EPR3}. Hence they supplement the measurement basis to define definite measurement outcomes. We therefore refer to these phases as `contextual phases'. The phases do not affect the theory until we make a measurement on separated subsystems.  

Before explaining why we wrote these two equivalent forms in Eqs. 2.1–2.4, with different classes of contextual phases, let's make an observation. Assume that the contextual phases are randomly installed into $\Psi_-$ sometime before we perform measurements with different polarizer settings. Overall there are eight phase combinations if we allow the phases to be $\pm 1$ only), but just two distinguished classes. We assume that half of the photon pairs will have class 1 contextual phases, half class 2 (see also Sec. 8 of Ref \cite{EPR3}). 

Imagine we could filter the photon pairs so that we only measure entangled photon pairs that have class 1 contextual phases. According to Proposal 2.1, by measuring the separated subsystem A, we acquire an observable for the vector $\frac{1}{\sqrt{2}}(|V'\rangle + |H'\rangle) \in \mathcal{H}_A$, which gives the observable H after collapse  during measurement. On the other hand, the vector $\frac{1}{\sqrt{2}}(|H'\rangle + |-V'\rangle) \in \mathcal{H}_B$ gives the observable V after collapse during measurement.  Therefore, if our measurement setup has the polarizers set at H for photon A and V for photon B (i.e. $\sigma_z \otimes \sigma_z$ basis detection), then \emph{every} pre-selected pair will pass through the polarizers and the polarizations are obviously anticorrelated.  

Conversely, if our imaginary filter selected only photon pairs that have class 2 contextual phases, and we set the measurement polarizers opposite to the previous scenario, now \emph{every} pre-selected pair will again pass through the polarizers because we will always find photon A to be V-polarized and photon B to be H-polarized. We can perform similar experiments in the $\sigma_x \otimes \sigma_x$ basis by setting the polarizers to $H'$ and $V'$. Then we see also the expected anticorrelation in the case of $\Psi_-$. These observations evidence the principle of separated subsystems.

The key result is that the measurement outcomes on each subsystem collapses the wavefunctions in the Hilbert spaces of the separated subsystems in an orchestrated fashion, without requiring interactions of any kind. That is, the contextual phases explain why measurement outcomes for the separated subsystems are correlated.

These results are consistent with the outcome of taking a partial trace of the density matrix of the pure state $|\Psi_-\rangle \langle \Psi_-|$ over subsystem B, which, as mentioned earlier, gives a mixed state for the measurement of subsystem A. That is, the measurement should be either eigenvalue H or V, with equal probability. 

Furthermore, these phases and their implications for correlations embedded in an entangled state evade detection by Bell-type inequalites because the underlying details of the contextual phases are invisible in the entangled state that exists in the tensor product of those Hilbert spaces (see the latter half of Sec. 8 in Ref \cite{EPR3}). Thus the contextual phases do not change any results or predictions of quantum states, except for allowing a mechanistic explanation of collapse of the state vector in the Hilbert space of a separated subsystem when it is measured. Note that the measurement outcomes are deterministic for a given contextual phase class, but the phase class of any $\Psi$ is random. The contextual phases cannot be detected in the Hilbert space of the composite system. We now give the technical basis for the contextual phases.

\section{Contextual phases in entangled states}

Each version of $\Psi_-$ in Eqs. 1.1, 2.1--2.4 is equivalent in $\mathcal{H}_A \otimes \mathcal{H}_B$. We can, as usual, predict measurements in the frame defined by the operator $\sigma_z \otimes \sigma_z$ (use Eqs 2.1, 2.2) or $\sigma_x \otimes \sigma_x$ (use Eqs 2.3, 2.4). That is sufficient to show that detection angles can be chosen to that Bell-type inequalities are violated (see, for example, Chapter 6 of \cite{Peres}).

If the state $\Psi_-$ is unchanged and the measurement correlations can violate Bell's inequality, how can the outcomes of measurements of the separated subsystems appear to be statistically pre-programmed? 

Owing to the wave basis for quantum mechanics, phase is an important factor in the construction of wavefunctions\cite{Dirac}. In terms of vectors in the Hilbert space of subsystem A, phase is encoded in the coefficients that define any normalized state vector with respect to our chosen basis, $\psi_A = \alpha|0\rangle_A + \beta|1\rangle_A$. Here we write the basis vectors as $|0\rangle$ and $|1\rangle$ for generality.  It is convenient to visualize the pair of complex coefficients $\alpha, \beta$ as a point on the sphere in four-dimensional real space, $\mathbb{S}^3$. Considering how observables are measured in quantum mechanics, as expectation values of the relevant operator, the overall (global) phase of the wavefunction is arbitrary. That is, points on $\mathbb{S}^3$ are identified by an equivalence relation $(\alpha, \beta) \sim (e^{i\phi}\alpha, e^{i\phi}\beta)$. For example, taking the global phase $\phi = \pi$, then
\begin{align*}
	\psi_A &= \alpha|0\rangle_A - \beta|1\rangle_A \sim -\alpha|0\rangle_A + \beta|1\rangle_A \\
	&= -(\alpha|0\rangle_A - \beta|1\rangle_A).
\end{align*}
In two-dimensions (e.g. force $\alpha$ and $\beta$ to be real) we see this equivalence relation identifies antipodal points on a circle ($\mathbb{S}^1$), Fig. 2a. 

\begin{figure}
	\includegraphics[width=6.5 cm]{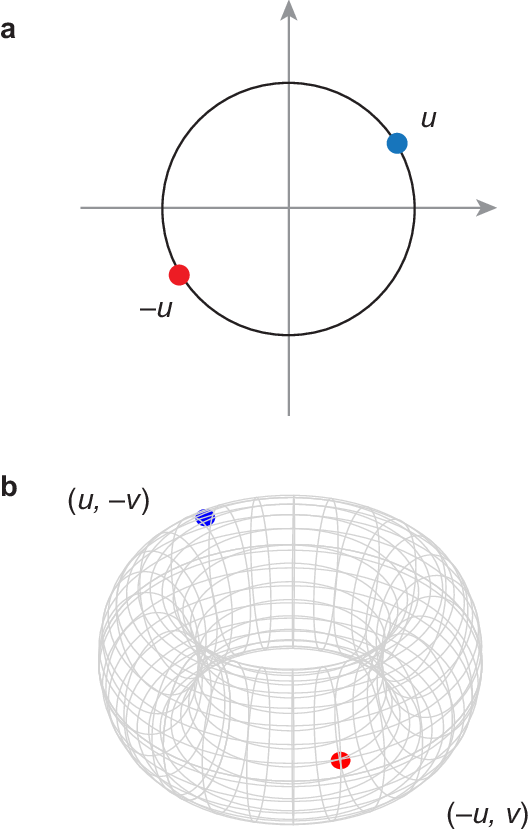}
	\caption{Geometric view of phase in wavefunctions. a. The coefficients of normalized state vectors $\psi_A = \alpha|0\rangle_A + \beta|1\rangle_A$ that differ by a global phase are antipodal points on the unit circle, $(\alpha, \beta) \sim -(\alpha, \beta)$. b. Two distinct points in $\mathbb{S}_A^1 \times \mathbb{S}_B^1$ that give equivalent vectors $-\psi_A \otimes \psi_B$ in the corresponding tensor product space. The red point is $(-\alpha_A, -\beta_A , \alpha_B, \beta_B)$ and the blue point is $(\alpha_A, \beta_A , -\alpha_B, -\beta_B)$. }
	\label{fig2}
\end{figure}  

Composite quantum states are formed by superpositions of the tensor product states $\psi_A \otimes \psi_B$. These include nonseparable (entangled) states such as $\Psi_-$. Like the states of the subsystems, global phase can be factored out of these states. The remaining, nonfactorizable, phase is that within the superpositions, for example the minus sign in Eq. 1.1. Elucidating how this phase is encoded in the subsystems, and the consequent implications for state vector collapse when measurements are made in the local Hilbert spaces (i.e. on separated subsystems), is the key result of the paper.  

Let's write a generalized version of Eq 1.1 as
\begin{equation*}
	\Psi = \phi_A \otimes \phi_B - \psi_A \otimes \psi_B .
\end{equation*}
Linearity of the tensor product implies that the second term in the superposition, $-(\psi_A \otimes \psi_B)$, is equivalent to $\psi_A \otimes (-\psi_B)$ and to $(-\psi_A) \otimes \psi_B$. However, lifting this vector to $\mathbb{S}_A^3 \times \mathbb{S}_B^3$, where we see the complex coefficients of $\psi_A$ and $\psi_B$ explicitly, it is clear that bilinear maps identify two \emph{distinct} points in $\mathbb{S}_A^3 \times \mathbb{S}_B^3$, for instance $(-\alpha_A, -\beta_A , \alpha_B, \beta_B)$ and $(\alpha_A, \beta_A , -\alpha_B, -\beta_B)$, Fig. 2b. Thus, in the tensor product space, the second term in Eq. 1.1 can take the different forms that account for how it maps from two possible points in $\mathbb{S}_A^3 \times \mathbb{S}_B^3$:
\begin{equation*}
	-(\psi_A \otimes \psi_B) = \psi_A \otimes (-\psi_B) = (-\psi_A) \otimes \psi_B .
\end{equation*}
The vector $\psi_A \otimes (-\psi_B)$ indicates that the phase in the superposition is explicitly encoded in the point on $\mathbb{S}^3_B$, while $(-\psi_A) \otimes \psi_B$ indicates that the phase in the superposition is explicitly encoded in a point on $\mathbb{S}^3_A$. The way phase in the superposition is identified with subsystem A or B is irrelevant for the composite states. However, the identification gives distinguished points on $\mathbb{S}_A^3 \times \mathbb{S}_B^3$. In other words, the same vector $-(\psi_A \otimes \psi_B)$ in the tensor product space is associated with two points on $\mathbb{S}_A^3 \times \mathbb{S}_B^3$, the points $( \psi_A , -\psi_B )$ and $(-\psi_A ,\psi_B)$. That matters when the two subsystems are separated because it controls measurement outcomes, as seen already for the example of $\Psi_-$. We focus on the points in $\mathbb{S}_A^3 \times \mathbb{S}_B^3$ because the Cartesian product supports canonical projection maps giving us the points on $\mathbb{S}^3_A$ and $\mathbb{S}^3_B$ (i.e. normalized vectors in $\mathcal{H}_A$ and $\mathcal{H}_B$ when the subsystems are separated).

Thus, for an \emph{entangled} state like $\Psi_-$, the phase of the superposition comes from either of two inequivalent combinations of complex coefficients for the subsystems, visualized as two inequivalent points in $\mathbb{S}_A^3 \times \mathbb{S}_B^3$. An antisymmetric state can be written $\Psi_a = x \otimes y + (-u)\otimes v$ or $\Psi'_a = x \otimes y + u \otimes (-v)$. A symmetric state can be written $\Psi_s = x \otimes y + u \otimes v$ or $\Psi'_s = x \otimes y + (-u) \otimes (-v)$. Like the global phase, these different underlying constructions of the tensor product states have no influence on observables. However, if we physically separate the subsystems from each other, then it turns out that the specific way the phase in the superposition is encoded in the subsystems \emph{does} matter because it controls how the state vector appears to collapse when measured. 

In hindsight, a way to realize that the association of the phase of a superposition to specific subsystems, which is essentially what is happening in our theory, may be more familiar than it first seems, is to consider phase flip gates. In the context of quantum operations, a phase flip gate changes the phase of one qubit, but the effect is can change an entangled state. For example, we can convert $\Phi_+$ to $\Phi_-$ by applying a phase flip gate to \emph{either} qubit A \emph{or} qubit B. This is analogous to our definition of class 1 and class 2 contextual phase. 

Contextual phase directs wavefunction collapse and thus explains why the measurement outcomes on separated subsystems are correlated. The contextual phase is not detectable, nor is it a classical local property of the state, and therefore it is not a local hidden variable. 

\section{Technical basis for contextual phases}

The conceptual basis of the interpretation of measurements proposed in this work is that the separated subsystems act to define ``axes'',  onto which measurements of an entangled state are projected. The axes are $\mathcal{H}_A$ and $\mathcal{H}_B$. This kind of canonical projection of the state vector cannot be made in the tensor product basis because that space is defined by linearizing the bilinear maps of vectors in the corresponding (algebraically covering) free vector space, which absorbs the bases of $\mathcal{H}_A$ and $\mathcal{H}_B$ into a new basis. Instead, we should consider the sets of vectors that are collected into equivalence classes in the tensor product space. These are the vectors in $\mathbb{S}_A^3 \times \mathbb{S}_B^3$, that are the subset of normalized vectors (or points) in the free vector space $\mathcal{F}(\mathcal{H}_A \times \mathcal{H}_B)$. The tensor product space $\mathcal{T}(\mathcal{H}_A \otimes \mathcal{H}_B)$ is defined from the covering free vector space $\mathcal{F}(\mathcal{H}_A \times \mathcal{H}_B)$ by the quotient\cite{TensorSpaces, Roman}:
\begin{equation*}
	\mathcal{T}(\mathcal{H}_A \otimes \mathcal{H}_B) = \mathcal{F}(\mathcal{H}_A \times \mathcal{H}_B)/\mathcal{R}(\mathcal{H}_A \times \mathcal{H}_B)
\end{equation*}
where $\mathcal{R}(\mathcal{H}_A \times \mathcal{H}_B)$ is the subspace with elements defined by 
\begin{eqnarray*}
	(\kappa v, w) - \kappa (v,w) \\
	(v, \kappa w) -  \kappa (v,w) \\
	(v_1 + v_2, w) - (v_1, w) - (v_2, w) \\
	(v, w_1 + w_2) - (v, w_1) - (v, w_2)
\end{eqnarray*}
with $\kappa \in \mathbb{K}$ (the field over which the vector space is defined) and we have $(v,w) \in \mathcal{F}(\mathcal{H}_A \times \mathcal{H}_B)$.  We can view these expressions as indicating equivalence relations. The notation used here, $\mathcal{F}(\mathcal{H}_A \times \mathcal{H}_B)$, is not entirely rigorous, but it is intended to convey that the formal linear combinations in $\mathcal{F}$ can be related to vector linear combinations in $\mathcal{H}_A$ and $\mathcal{H}_B$. Details are developed in another paper\cite{sepsubs}.

The intuition for a quotient space is that we ``group'' elements together according to an equivalence relation, then we treat each group (i.e. each equivalence class) as a single point in the quotient space. For example in $\mathbb{Z}/\mathbb{Z}_3$, the (infinite) set of integers $\{ \dots -6,-3, 0, 3, 6, \dots \}$ is denoted by the equivalence class [0], $\{ \dots -5, -2, 1, 4, 7, \dots \}$ by [1], and $\{ \dots -4,-1, 2, 5,8, \dots \}$ by [2]. Knowing the element in $\mathbb{Z}/\mathbb{Z}_3$ (i.e., 0, 1, or 2) does not indicate a unique corresponding element in $\mathbb{Z}$, but the reverse is true. When we form a quotient space or set $\mathcal{F}/\mathcal{R}$, we do not remove the elements from $\mathcal{F}$, instead we group them into equivalence classes.

Consider another example,  $\mathbb{Z}/\{\pm 1\}$, which is the set of equivalence classes under the relation $n \sim m$ if and only if $m = \pm n$. Therefore $[n] = \{-n, n\}$ for $n \ne 0$ and $[0] = \{0\}$. Let's say we are given $[3]$. We cannot know with whether this is denotes specifically $-3$ or $3$, because it indicates the set $\{-3, 3\}$.  However, if we perform a suitable measurement on one item denoted $[3]$, we could resolve whether the item is $-3$ or $3$, either being equally likely. 

That example clarifies the distinction between an entangled state $\Psi$ and the states being measured via the separated subsystems. $\Psi \in \mathcal{H}_A \otimes \mathcal{H}_B$ is an equivalence class of vectors in $\mathcal{F}(\mathcal{H}_A \times \mathcal{H}_B)$. Therefore $[\Psi] = (\phi_A, \phi_B) + \kappa (\psi_A, \psi_B)$ refers to elements in a set that we partition, according to the two possible measurement outcomes, as $\{ \{\text{class 1}\}, \{\text{class 2}\} \}$. In terms of the maximally entangled states, with $\kappa = -1$, $\{\text{class 1}\}$ has the representative element $\Psi_{\text{as}}(\text{class 1}) = (\phi_A, \phi_B) + (\psi_A, -\psi_B)$. While a representative element of $\{\text{class 2}\}$ is $\Psi_{\text{as}}(\text{class 2}) = (\phi_A, \phi_B) + (-\psi_A, \psi_B)$. Whereas, with $\kappa = +1$, $\{\text{class 1}\}$ has the representative element $\Psi_{\text{s}}(\text{class 1}) = (\phi_A, \phi_B) + (\psi_A, \psi_B)$ and $\{\text{class 2}\}$ has the representative element $\Psi_{\text{s}}(\text{class 2}) = (\phi_A, \phi_B) + (-\psi_A, -\psi_B)$. In this notation, `as' means antisymmetric and `s' means symmetric. 

This changes the view of collapse of the wavefunction under measurement from being a probabilistic event to it being a statistical property of outcomes of measurements on vectors identified in a quotient space. Given $\Psi \in \mathcal{T}(\mathcal{H}_A \otimes \mathcal{H}_B)$, we have a class $[\Psi] \in \mathcal{F}(\mathcal{H}_A \times \mathcal{H}_B)$ that comprises the sets $\{\text{class 1}\}$  and $\{\text{class 2}\}$, differentiated by contextual phases. Whether a measurement reveals an element from $\{\text{class 1}\}$  or from $\{\text{class 2}\}$ is assumed to be equally likely because there does not appear to be a reason why one configuration of contextual phase would be favoured over another.

We can also consider the interpretation in terms of topology of these state spaces. Consider the map from a point on $\mathbb{S}_A^3 \times \mathbb{S}_B^3$ to normalized simple tensors:
\begin{equation}
	\mathbb{S}_A^3 \times \mathbb{S}_B^3 \rightarrow \mathcal{H}_A \otimes \mathcal{H}_B, \quad (u,v) \mapsto u \otimes v.
\end{equation}

For real spaces we only need to make the identification in $\mathcal{H}_A \otimes \mathcal{H}_B$
\begin{equation*}
	(u,v) \sim (-u, -v),
\end{equation*}
so the image of the quotient map is homeomorphic to
\begin{equation}
	M = (\mathbb{S}_A^3 \times \mathbb{S}_B^3)/\{ (u,v) \sim (-u, -v) \},
\end{equation}
which has a fundamental group $\mathbb{Z}_2$. $M$ has the same dimension as the manifold $\mathbb{S}_A^3 \times \mathbb{S}_B^3$ and is, similarly, a smooth, compact, connected 6-manifold. However, the difference in fundamental group means they are not homeomorphic. 

The quotient is two-to-one, so every simple tensor in $\mathcal{H}_A \otimes \mathcal{H}_B$ has two representatives in $\mathbb{S}_A^3 \times \mathbb{S}_B^3$, namely $(u,v)$ and $(-u, -v)$. This is the reason that there are two contextual phase classes. From a topological perspective, the quotient map is a two-sheeted covering. That means that, locally, it looks like two separate copies of the target lying over each small neighbourhood. In other words, at each point on the manifold $M$ we have the two pre-images $(u,v)$ and $(-u, -v)$.

\section{Nonlocal correlations}

In the prior sections it was established that the contextual phases give the information that is lost when the Cartesian product $\mathbb{S}_A^3 \times \mathbb{S}_B^3$ is linearized by the bilinear map to the tensor product space. That contextual information allows the subsystems to be separated in the sense that we can recover the vectors in $\mathcal{H}_A$ and $\mathcal{H}_B$ because they are given by the points in $\mathbb{S}_A^3$ and $\mathbb{S}_B^3$, isolated by the canonical map.  In particular, the analysis provided above suggests how correlated measurement outcomes for two entangled subsystems are explicitly embedded in an entangled state by the contextual phases. 

\emph{State collapse for local measurement of isolated subsystems.} Collapse of a superposition local to an isolated subsystem A associated with an entangled state in $\mathcal{H}_A \otimes \mathcal{H}_B$,  that is, a superposition of vectors where each is an element of $\mathcal{H}_A$, can be reinterpreted as interference of a superposition of state vectors in the Hilbert space $\mathcal{H}_A$ caused by measurement of the isolated subsystem. According to the contextual phases randomly encoded in the state, the interference yields a measurement outcome in the chosen measurement basis. That is, while the contextual phases are random, the collapse induced by measurement is correlated among the separated systems.

The foundation of this hypothesis is to define measurement of the isolated subsystems of an entangled state using the principle of separated subsystems (Proposal 2.1). That proposal, which is consistent with the proposal by von Neumann that measurements of separated subsystems should be obtained by observables of vectors in the subsystem Hilbert spaces\cite{vonNeumann}, was justified in the present work by consistency arguments. It was explained further in the previous sections that we can think about an entangled state in $\mathcal{H}_A \otimes \mathcal{H}_B$ as a superposition of points in $\mathbb{S}_A^3 \times \mathbb{S}_B^3$, so that the contextual phases are explicit. We might describe that as a lift from the tensor product space to the free vector space, but really it is not a map---it is a realization via measurements on the separated subsystems of elements identified by their equivalence class in the tensor product space. 

Once we have identified vectors in the free vector space, which are defined by the Cartesian product of basis vectors, then the canonical map provides the superpositions of vectors in $\mathcal{H}_A$ and $\mathcal{H}_B$. Given these local superpositions, straightforward computation shows the inevitable correlations between the outcomes of measurements when the superposition in each local Hilbert space collapses. The significant outcome is that an \emph{ad hoc} random collapse does not need to be invoked because the collapse is a natural consequence of the interference encoded by the contextual phase. The proof is completed by exhibiting the remaining three maximally-entangled bipartite states. 

Note that usually the $x$-basis form of $\Phi_-$ is labeled $\Psi_+$ and, conversely, the $x$-basis form of $\Psi_+$ is labeled $\Phi_-$. Here we do not change the label. Instead we keep the $z$-basis notation regardless of the basis in order to emphasize that these wavefunctions are identical.

In the $x-$basis we have
\begin{align}
	\Phi_-(\text{class 1}) &= \frac{1}{\sqrt{2}} \Big[ |V'\rangle_A \otimes |H'\rangle_B + |H'\rangle_A \otimes |V'\rangle_B  \Big] \\
	\Phi_-(\text{class 2}) &= \frac{1}{\sqrt{2}} \Big[ |V'\rangle_A \otimes |H'\rangle_B + |-H'\rangle_A \otimes |-V'\rangle_B  \Big] ,
\end{align}
\begin{align}
	\Phi_+(\text{class 1}) &= \frac{1}{\sqrt{2}} \Big[ |H'\rangle_A \otimes |H'\rangle_B + |V'\rangle_A \otimes |V'\rangle_B  \Big] \\
	\Phi_+(\text{class 2}) &= \frac{1}{\sqrt{2}} \Big[ |H'\rangle_A \otimes |H'\rangle_B + |-V'\rangle_A \otimes |-V'\rangle_B  \Big] ,
\end{align}
\begin{align}
	\Psi_+(\text{class 1}) &= \frac{1}{\sqrt{2}} \Big[ |H'\rangle_A \otimes |H'\rangle_B + |V'\rangle_A \otimes |-V'\rangle_B  \Big] \\
	\Psi_+(\text{class 2}) &= \frac{1}{\sqrt{2}} \Big[ |H'\rangle_A \otimes |H'\rangle_B + |-V'\rangle_A \otimes |V'\rangle_B  \Big] .
\end{align}
In the $z-$basis these states are
\begin{align}
	\Phi_-(\text{class 1}) &= \frac{1}{\sqrt{2}} \Big[ |H\rangle_A \otimes |H\rangle_B + |-V\rangle_A \otimes |V\rangle_B  \Big] \\
	\Phi_-(\text{class 2}) &= \frac{1}{\sqrt{2}} \Big[ |H\rangle_A \otimes |H\rangle_B + |V\rangle_A \otimes |-V\rangle_B  \Big] ,
\end{align}
\begin{align}
	\Phi_+(\text{class 1}) &= \frac{1}{\sqrt{2}} \Big[ |H\rangle_A \otimes |H\rangle_B + |V\rangle_A \otimes |V\rangle_B  \Big] \\
	\Phi_+(\text{class 2}) &= \frac{1}{\sqrt{2}} \Big[ |H\rangle_A \otimes |H\rangle_B + |-V\rangle_A \otimes |-V\rangle_B  \Big] ,
\end{align}
\begin{align}
	\Psi_+(\text{class 1}) &= \frac{1}{\sqrt{2}} \Big[ |H\rangle_A \otimes |V\rangle_B + |V\rangle_A \otimes |H\rangle_B  \Big] \\
	\Psi_+(\text{class 2}) &= \frac{1}{\sqrt{2}} \Big[ |H\rangle_A \otimes |V\rangle_B + |-V\rangle_A \otimes |-H\rangle_B  \Big] .
\end{align}

We compare in Table 1 measurement outcomes for wavefunctions incorporating either class 1 or class 2 contextual phases predicted by collapse of the superpositions in $\mathcal{H}_A$ and $\mathcal{H}_B$. The results come from inspection of Eqs. 2.1--2.4 and 5.1--5.12. The outcomes can be thought of as the polarizer settings for which all separated entangled photon pairs of a class of contextual phases are transmitted.  The expectation values of $\langle \Psi | \sigma_i \otimes \sigma_i | \Psi \rangle$ are also listed for the two measurement directions (they are identical for class 1 and class 2 contextual phases).   Notice that the class 1 and class 2 measurement outcomes are indistinguishable by permutation of the labels A and B. That is, the measurements cannot reveal whether the state has class 1 or class 2 contextual phase.

\begin{table}[!h]
	\caption{Correlations for measurements of separated subsystems in various entangled states with different detector settings\textsuperscript{a}}
	\label{table_1}
	\begin{tabular}{llll}%%%The number of columns has to be defined here
		\hline
		State &    & $\sigma_z \otimes \sigma_z$ & $\sigma_x \otimes \sigma_x$ \\
		\hline
		$\Psi_-$    &  expectation values  & $-1$  & $-1$   \\
		&  class 1 outcomes  & $H$, $V$  & $V'$, $H'$ \\
		&  class 2 outcomes  & $V$, $H$  & $H'$, $V'$ \\
		$\Psi_+$    &  expectation values  & $-1$   & $+1$  \\
		&  class 1 outcomes  & $H$, $V$  & $H'$, $H'$ \\
		&  class 2 outcomes  & $V$, $H$  & $V'$, $V'$ \\
		$\Phi_-$    &   expectation values  & $+1$  & $-1$  \\
		&  class 1 outcomes  & $H$, $H$  & $V'$, $H'$ \\
		&  class 2 outcomes  & $V$, $V$  & $H'$, $V'$ \\
		$\Phi_+$    &   expectation values  & $+1$  & $+1$  \\
		&  class 1 outcomes  & $H$, $H$  & $H'$, $H'$ \\
		&  class 2 outcomes  & $V$, $V$  & $V'$, $V'$ \\ 
		\hline
	\end{tabular}
	\footnote{\textsuperscript{a} Usually the $x$-basis form of $\Phi_-$ is labeled $\Psi_+$ and, conversely, the $x$-basis form of $\Psi_+$ is labeled $\Phi_-$. Here we do not change the label. Instead we keep the $z$-basis notation regardless of the basis in order to emphasize that these wavefunctions are identical.}
	\vspace*{-4pt}
\end{table}

The following hypothesis explains how the sequence of measurements on one subsystem give random outcomes, because the sequence is random with respect to whether the phase of the contextual phase of the wavefunction is class 1 or 2. Whereas, the measurement outcomes on the two subsystems are correlated, because the contextual phase is correlated across the subsystems. 

\emph{Measurement statistics are determined by the random distribution of contextual phase classes in an ensemble of entangled states.} To support this hypothesis, assume that the appropriate contextual phases are randomly fixed in an entangled state prior to measurement. That could happen when the state is prepared, or as the subsystems separate. It follows that a sequence of measurements of the state of either subsystem will give random outcomes because the contextual phases are random. The proposition also follows as a consequence of our interpretation of the wavefunction in $\mathcal{H}_A \otimes \mathcal{H}_B$ as an equivalaence class of vectors in the corresponding free vector space from which the tensor product space derives. As a consequence of the correlations \emph{between} the contextual phases in each subsystem, joint measurements of the subsystems give correlated or anticorrelated observables, according to the expectation value for the operator $\sigma_i \otimes \sigma_i$ acting on the state, where $\sigma_i$ is any of the three Pauli operators.

\section{Summary and physical examples}

When a composite system is in an entangled state, the subsystems are locked in superposition. Measurements of this state are described by expectation values of operators on vectors in the Hilbert space $\mathcal{H}_A \otimes \mathcal{H}_B$. A fascinating aspect of quantum systems is that their physical building blocks---photons, electrons, atoms, and so on---can be separated in space. This means that we also need to describe measurements on the separated subsystems. As von Neumann indicates in Chapter VI.2 of \cite{vonNeumann}, such measurement outcomes correspond to the expectation values in the Hilbert spaces of the separated subsystems, $\mathcal{H}_A$ or $\mathcal{H}_B$. However, $\mathcal{H}_A$, $\mathcal{H}_B$, and $\mathcal{H}_A \otimes \mathcal{H}_B$ are all different Hilbert spaces and it is not obvious how to, for example, extract a vector $v_A \in \mathcal{H}_A$ given a vector $v_{AB} \in \mathcal{H}_A \otimes \mathcal{H}_B$, except when $v_{AB}$ is a separable state. Indeed, only in the case of separable states do we find that eigenvalues $\lambda_i$ for some operator $T_A$ and vectors $x_i^A \in  \mathcal{H}_A$:
\begin{equation*}
	T_A x_i^A = \lambda_i x_i^A,
\end{equation*}
are also eigenvalues for the related operator on states in $\mathcal{H}_A \otimes \mathcal{H}_B$:
\begin{equation*}
	(T_A \otimes \mathbb{I}_B) (x_i^A \otimes x_j^B) = \lambda_i (x_i^A \otimes x_j^B),
\end{equation*}
where $\mathbb{I}_B$ is the identity operator on subsystem B. 

The present work proposes a way to deal with this issue by identifying entangled states in $\mathcal{H}_A \otimes \mathcal{H}_B$ to states that represent the appropriate local superpositions in the separated, non-interacting subsystems. Those states, in accord with von Neumann's proposal, lie in $\mathcal{H}_A$ and $\mathcal{H}_B$. The identification we have proposed in the present paper associates the states in $\mathcal{H}_A \otimes \mathcal{H}_B$ to a superposition of vectors in the covering free vector space. Vectors associated with each separable state component of the entangled state are normalized in their respective Hilbert space---hence, these vectors are the vectors in $\mathbb{S}^3_A \times \mathbb{S}^3_B$. The important consideration here is to account for the possible ways that the phase of the superposition lifts to these vectors, Sec. 3. The various points in $\mathbb{S}^3_A \times \mathbb{S}^3_B$ that map to a single vector in $\mathcal{H}_A \otimes \mathcal{H}_B$ are indicated as classes of state vectors with different contextual phase factors. 

Having defined vectors in $\mathcal{H}_A$ and $\mathcal{H}_B$ corresponding to the separated subsystems, we then consider measurement outcomes. These outcomes depend on the measurement performed---for instance, whether the polarizer is set to $H$ or $H'$. We have shown for any measurement there is a way that the state vector of the separated subsystem will collapse, without making any \emph{ad hoc} assumptions, to yield a specific measurement outcome. This captures the fact that a quantum system is particle-like, so there must be only one outcome for each measurement, for example the photon is transmitted through a polarizer, or it is not. That outcome is enabled by the wave-like properties of the particle, which allow suitable interference of the superposition of vectors in the separated subsystem's Hilbert space. Measurement outcomes depend on the contextual phases, as well as the way these phases encode correlations between the the subsystems (see Table 1).

Returning to the example of the beam splitter, it is known\cite{KnightQOptics} that the measurement outcomes---whether the incident photon is reflected through `port' A or transmitted through port B---come from the entangled state
\begin{equation}
	\Psi_{BS} = \frac{1}{\sqrt{2}} \Big( |0\rangle_A |1\rangle_B + i|1\rangle_A |0\rangle_B\Big).
\end{equation}
The probabilistic outcomes predicted from this state is that a photon is either transmitted or reflected, as expected, and as found by experiments\cite{Grangier1986}. Considering the relevant contextual phases, we can predict these outcomes at the state vector level (not solely based on the density matrix). For example, two of the equivalent ways of writing $\Psi_{BS}$ are
\begin{align}
	\Psi_{BS}(\text{class 1}) &= \frac{1}{\sqrt{2}}\Big[ \frac{1}{\sqrt{2}}(|0\rangle_A + |1\rangle_A) \otimes \frac{1}{\sqrt{2}}(i|0\rangle_B + |1\rangle_B) \\
	&+ \frac{1}{\sqrt{2}}(|0\rangle_A - |1\rangle_A) \otimes \frac{1}{\sqrt{2}}(-i|0\rangle_B + |1\rangle_B) \Big] \nonumber
\end{align}
\begin{align}
	\Psi_{BS}(\text{class 2}) &= \frac{1}{\sqrt{2}}\Big[ \frac{1}{\sqrt{2}}(|0\rangle_A + |1\rangle_A) \otimes \frac{1}{\sqrt{2}}(i|0\rangle_B + |1\rangle_B) \\
	&+ \frac{1}{\sqrt{2}}(-|0\rangle_A + |1\rangle_A) \otimes \frac{1}{\sqrt{2}}(i|0\rangle_B - |1\rangle_B) \Big] \nonumber
\end{align}

Expanding term-by-term, we see that Eqs. 6.1--6.3 are identical. The class of contextual phase is hypothesized to be installed jointly in the separated subsystems randomly before they are separated. 

Notice how class 1 contextual phase gives the local state, in $\mathcal{H}_A$, 
\begin{equation*}
	\psi_A = \frac{1}{2}(|0\rangle_A + |1\rangle_A) + \frac{1}{2}(|0\rangle_A - |1\rangle_A).
\end{equation*}
Upon measurement, $\psi_A$ collapses to $|0\rangle_A$; predicting that no photon is reflected. For the associated local state in $\mathcal{H}_B$, we have 
\begin{equation*}
	\psi_B = \frac{1}{2}(i|0\rangle_B + |1\rangle_B) - \frac{1}{2}(-i|0\rangle_B + |1\rangle_B),
\end{equation*}
that collapses, when measured, to give $|1\rangle_B$; a photon is transmitted. Class 2 contextual phase gives the opposite outcomes. 

These measurement outcomes are a natural consequence of measurements on the separated subsystems and do not require an \emph{ad hoc} notion of collapse. The measurement value reflects the property of a photon (which means a photon cannot be partially transmitted or reflected), and such a physical outcome is enabled by the way wave-like superpositions interfere during measurement of a separated subsystem. Importantly, the collapse of the wavefunction of each separated subsystem is perfectly correlated by the contextual phase. So that if no photon is reflected, then a photon is definitely transmitted. 

A second example can also be experimentally verified. Furthermore, this example compares and contrasts measurements on states of the composite system to those of separated subsystems. Consider two interacting atoms or molecules, labelled A and B, in the first excited electronic state. The excited states of the composite system are Frenkel or molecular excitons\cite{Bardeen2014, Mirkovic2017}. Let's say the lowest lying electronic state, an element of $\mathcal{H}_A \otimes \mathcal{H}_B$ is
\begin{equation*}
	\Psi_+ = \frac{1}{\sqrt{2}} \Big( |0\rangle_A |1 \rangle_B + |1\rangle_A |0 \rangle_B \Big),
\end{equation*}
where 0 denotes the ground state and 1 means an electronic excited state in the site (computational) basis. We can calculate the transition dipole strength for fluorescence emission from $\Psi_+$ to the ground state\cite{Mirkovic2017}:
\begin{align*}
	|\mu_+|^2 &= | \langle \Psi_+ | \boldsymbol{\mu} | \Psi_0  \rangle |^2 \\
	&= 2|\mu_0|^2,
\end{align*}
where $\Psi_0$ is the ground state wavefunction of the dimer, $\boldsymbol{\mu}$ is the dipole operator, and $|\mu_0|$ is the magnitude of the transition dipole moment for either subsystem in isolation (they are assumed to be equal), We further assume the system is completely isolated from the environment. The enhancement of the transition dipole strength owing to the collective emission of the atom or molecular pair is a form of superradiance. 

Now consider that we generate the exciton state $\Psi_+$, then quickly separate the atoms/molecules so that the interaction between them becomes negligible---yet the isolated composite system remains in the entangled state. Now let's observe only subsystem A and measure whether or not it emits a photon by fluorescence. To obtain a measurement outcome, the conventional approach is to say the wavefunction of the subsystem randomly collapses to give either the ground or excited state of A. Aside from recognition that the concept of random collapse is arbitrary, it is unclear how correlations between the subsystems arise so that if one atom is excited, the other must be in its ground state. From the perspective of the contextual phase model these concerns are resolved.

The results of measurements on the separated subsystems, Table 1, will be in accord with physical expectations without needing to invoke a random collapse phenomenon. Moreover, if we gate the measurements of emitted photons for an ensemble of measurements to time-resolve the fluorescence, then we expect to find that the radiative rate is equal to that of one molecule in isolation---it cannot be superradiatively enhanced when the atoms/molecules are sufficiently separated. That is precisely the prediction of the contextual phase model, which gives half the ensemble of A subsystems to be in the ground state and half to be in the excited state, with an opposite correlation for subsystem B. 

Thus the contextual phases correctly predict the different measurement outcomes on the composite system compared to measurements on the separated subsystems.

\section{More than two subsystems}

% GHZ example

The examples can  be extended to systems comprising greater than two subsystems. To illustrate that, we consider now the GHZ state\cite{GHSZ1990}. Experiments on the GHZ state comprising three entangled photons have allowed compelling tests of `local realism' by showing a clear contradiction between expectations of local realism and quantum mechanics\cite{Zeilinger2000}. This conclusion results from considering how the GHZ state in the Hilbert space $\mathcal{H}_A \otimes \mathcal{H}_B \otimes \mathcal{H}_C$ comprises certain possible polarization sequences. Now it will be shown that, while indeed local realism is contradicted by quantum mechanics (none of the prior results change), the principle of separated subsystems does hold. 

To examine this principle, we should focus on measurements of the polarizations of separated photons in the local Hilbert spaces $\mathcal{H}_A$, $\mathcal{H}_B$, and  $\mathcal{H}_C$. As an example, write the $x-$basis form of $\Psi_{\text{GHZ}}$ in two representative ways that indicate class 1 and class 2 contextual phases:

\begin{align}
	\Psi_{\text{GHZ}} &= \frac{1}{\sqrt{2}}\Big(  |H\rangle_A |H\rangle_B |H\rangle_C + |V\rangle_A |V\rangle_B |V\rangle_C \Big) \\
	\begin{split}
		\Psi_{\text{GHZ}}(\text{class 1}) &=  \frac{1}{2}\Big(  |H'\rangle_A |H'\rangle_B |H'\rangle_C +  |H'\rangle_A |V'\rangle_B |V'\rangle_C  \\
		&+   |V'\rangle_A |H'\rangle_B |V'\rangle_C  +  |V'\rangle_A |V'\rangle_B |H'\rangle_C   \Big) 
	\end{split} \\
	\begin{split}
		\Psi_{\text{GHZ}}(\text{class 2}) &=  \frac{1}{2}\Big(  |H'\rangle_A |H'\rangle_B |H'\rangle_C +  |H'\rangle_A |-V'\rangle_B |-V'\rangle_C  \\
		&+   |-V'\rangle_A |H'\rangle_B |-V'\rangle_C  +  |-V' \rangle_A |-V'\rangle_B |H'\rangle_C   \Big) .
	\end{split}
\end{align}

Equivalent forms that give $\Psi_{\text{GHZ}}$ up to a global phase are obtained by choosing phases for the class 2 state as follows.  (i) For the first term choose for each subsystem an antipodal phase for each $| V \rangle$ term and a (possibly different) antipodal phase each $| H\rangle$ term. (ii) For successive terms, permute these phases. (iii) Ensure that every term contains either an even total number of antipodal vectors $ |-V'\rangle_A$ and $|-H'\rangle_A $, or every term contains an odd number of antipodal vectors. The latter condition yields a global phase of $-1$. 

Now consider measurements on the separated subsystems (photons in this example). Assume the source of GHZ-entangled photons randomly produces either class 1 or class 2 contextual phases. Set the measurement polarizers to all H. Half the incident photon triples will be class 1, and only these sets of photons will be transmitted through the polarizers because collapse of the superposition of state vectors in each local Hilbert space gives an H polarized photon in each isolated subspace. For example, consider photon A. When separated, its state vector in $\mathcal{H}_A$ is the superpositon $\frac{1}{2} (|H'\rangle_A + |H'\rangle_A + |V'\rangle_A + |V'\rangle_A )$, which collapses when measured to give a measured observable for $|H\rangle_A $. Contrast that to the class 2 contextual phases that cause the isolated state vector for photon A to collapse to $|V\rangle_A $. Finally, note that if we set the polarizers to measure any of the polarization combinations $H'H'H'$, $H'V'V'$, $V'H'V'$ or $V'V'H'$, as in the experiments reported by Pan and co-workers\cite{Zeilinger2000}, the three photons are transmitted, regardless of whether the entangled state has class 1 or class 2 contextual phases. 

That is, the equally probable correlated measurement outcomes predicted by quantum mechanics are encoded as the contextual phases that ensure that measurements of entangled subsystems show correlations without requiring any kind of interaction or communication during the independent measurements.

\section{Conclusion}

The specific advance of this paper is to propose how contextual phases could feasibly explain the mechanism of collapse when measurements are made on subsystems separated from an entangled state. It was hypothesized that nonlocal  contextual phases programmed randomly into the entangled state, either when the state is prepared or when the subsystems are separated, predetermine the measurement outcomes on the subsystems. The correlations observed, however, come from the entangled state. Thus, a sequence of measurements on either subsystem gives a random sequence of outcomes. However, joint measurements are correlated, evidenced by classical measurement outcomes that are integrated into entangled states by the contextual phases. 

What we have done is to change the point of view from the statistics of measurement outcomes coming from \emph{random collapse} of the superpositions---where nonlocal interactions are invoked to explain correlations in joint measurements---to the perspective of an ensemble of entangled states that are identical to each other except for a hidden random phase in the underlying superpositions. The phase is invisible to states in $\mathcal{H}_A \otimes \mathcal{H}_B$, but it is exposed by measurements on the separated subsystems.  The phases direct the collapse of each superposition---which is interpreted as an interference phenomenon---to a particular classical outcome when a subsystem is measured. The measuring apparatus thus obtains a classical read-out of the quantum correlations embedded in an entangled state. 

The new perspective offered by contextual phase is that the confusing problem of nonlocal correlations highlighted by EPR\cite{EPR} and subsequent work can be solved in the existing framework of quantum mechanics, while remaining, consistent with Bell's theorem and analyses of objective reality. Central to this conclusion is a mechanism of collapse proposed here, that, more broadly, will likely solidify the theory of measurement of quantum states.

%%%%%%%%%%%%%%%%%%%%%%%%%%%%%%%%%%%%%%%%%%

\enlargethispage{20pt}

%\ethics{Insert ethics text here.}

%\dataccess{There are no new data reported in this paper.}

%\aucontribute{Insert author contribute text here.}

\begin{acknowledgments}
	Dr. Garry Rumbles is thanked for discussions about the manuscript. Financial support was provided by the Division of Chemical Sciences, Geosciences and Biosciences, Office of Basic Energy Sciences,of the US Department of Energy through grant no. DE-SC0015429. 
\end{acknowledgments}

%%%%%%%%%%%%%%%%%%%%%%%%%%%%%%%%%%%%%%%%%%

%%%%%%%%%%%%%%%%%%%%%%%%%%%%%%%%%%%%%%%%%%
\vspace{6pt} 

%%%%%%%%%%%%%FOR ENTROPY

%\acknowledgments{}This research was funded by the National Science Foundation under Grant No. 2211326.}

%%%%%%%%%%%%

\bibliography{Scholes_bib_July2026}
\bibliographystyle{tfq}

\end{document}